\documentclass[a4paper,10pt]{article}
\title{ Fourth Quantization}
\author{ 
Mir Faizal\\ Mathematical Institute, University of Oxford
\\ Oxford
OX1 3LB, United Kingdom   }
\date{}
\begin{document}

\maketitle

\begin{abstract}
In this paper we will analyse the creation of the  multiverse. We will first calculate 
the wave function  for the multiverse using third quantization. Then we will fourth quantize
this theory. We 
will  show that there is no single vacuum state for this theory. Thus,
  we can end up with a multiverse, even after starting from a vacuum state.
 This will  be used as a possible explanation for the 
creation of the multiverse. 
We also analyse the effect of interactions in this fourth quantized theory. 
\end{abstract}
\section{Introduction}

The existence of the multiverse first appeared 
in the many-worlds interpretation
of quantum theory
\cite{1a}. This idea has now resurfaced in the 
landscape of string theory  \cite{3a, 4a}. 
As this landscape is populated by $10^{500}$ 
vacuum states \cite{1h}, the possibility of all of
 them being real vacuum states 
for different universes remains an open one \cite{1w}.   
As a matter of fact, this model of the multiverse 
has also been used as an explanation 
for inflation in  chaotic inflationary multiverse  \cite{2a,  q2, q3}. 
In fact, it is expected that the number of universes in the chaotic inflationary multiverse 
will be even more than the number of different string theory vacuum states. This is because 
 even for a  single string theory vacuum state,
the large-scale structure  and the matter content in each of the locally 
Friedman parts will  be different \cite{q1}. Thus, we should in principle be studding 
a collection of multiverse, each for a different string theory vacuum. 

It may be noted that the big bang can be explained as the collision
of two universes to form two new universes in the multiverse \cite{a,b}. 
Thus, to explain the cause of 
big bang, we require a theory of many universes, where these universes can interact with each other. 
Third quantization forms a natural formalism to study the such a model of the multiverse \cite{1,2}. 
This is because it is well known  that we cannot explain the creation and annihilation of particles using  a single particle 
quantum mechanics and we have to resort to a second quantized theory for that purpose. In the  second 
quantized formalism,   the Schroedinger 
equation is treated as a classical field equation and it is quantized one more time. 
  The creation and annihilation of particles is explained, by adding interaction terms to it,  
The Wheeler-DeWitt equation acts like the Schroedinger equation for quantum gravity \cite{tyu}. So, in analogy with the 
single particle quantum mechanics, we cannot explain the creation and annihilation of universes using a second quantized 
Wheeler-DeWitt equation, and we have to use a third quantized formalism for doing that purpose \cite{2ba,3ba}. In doing so 
the Wheeler-DeWitt equation is treated as a classical field equation and interactions are added to it. These interaction terms 
cause the universes to get created and annihilated. 
 The third quantization of the Brans-Dicke theories
  \cite{i} and Kaluza-Klein theories  \cite{ia} 
 have been already  thoroughly studied. Virtual black holes in third quantized $f(R)$ gravity has also been studied \cite{1111}. 
In doing so a solution to the problem of time has also been proposed. 
The uncertainty principle for third quantization of $f(R)$ gravity has also been analysed \cite{1q}.
If we go to fourth quantization we will go to a theory of multi-multiverses.  This seems to be a natural 
structure that will occur in chaotic inflationary multiverse because 
 there will be a multiverse for each different string theory vacuum state. 

In this paper we will analyse  the Wheeler-DeWitt equation in  minisuperspace approximation.
 We will first 
obtain an explicit expression for the wave function of the multiverse. Then we will analyse the creation of 
multiverse from a fourth  vacuum state. Finally, we will analyse the effect of interaction 
of different multiverses.  It may be noted just like in the third quantized formalism, the creation of universe can be 
explained, for the creation of the multiverse, we will have to quantize the third quantized theory one more time. 
Thus, in this paper fourth quantization is studied for the first time. 

\section{Third Quantization}
In the first quantized 
formalism, both the non-relativistic Schroedinger equation and
the relativistic  Klein-Gordon equation  are viewed as quantum mechanical equations. 
These equations describe the evolution of the  wave function for a single particle. 
However, when we go to the second quantized formalism, both these first quantized equations are no longer viewed 
as quantum mechanical equations. They are rather viewed as classical field equations and the
 first quantized wave function is viewed as a classical field, in the second quantized formalism.
A Lagrangian  can be constructed which generates these classical field equations. 
This Lagrangian is used to calculate the 
momentum conjugate to the classical field. After that  a second quantized equal time commutator 
is  imposed between the operator corresponding to this momentum variable and the operator 
corresponding to the field variable. This is how a second quantized the 
Fock space is constructed.  
 
The Wheeler-DeWitt equation can be viewed as the  Schroedinger's equation for a single universe. 
It is already a second quantized equation. However, just as we could go from  first quantization 
 to second quantization, by viewing the first quantized  equation as a classical field equation, 
we can go from  second quantization to third quantization, by viewing the Wheeler-DeWitt equation 
as a classical field equation. We can also construct a Lagrangian which generates this classical field equation. 
Furthermore, from this Lagrangian a momentum conjugate to this new field variable can be calculated. 
This momentum can be used to third quantize this theory, by imposing 
equal time commutator with this new field variable. This way we will obtain a Fock space for the multiverse. 

Now we will analyse third quantization of Wheeler-DeWitt equation in the minisuperspace approximation. Thus, we start with 
 the 
Friedman-Robertson-Walker metric, which is given by 
\begin{equation}
ds^{2}=-N^{2}dt^{2}+a^{2}\left(  t\right)  d\Omega_{3}^{2},
\end{equation}
where $d\Omega_{3}^{2}$ is the usual line element on the three sphere and $N$ is the
lapse function and here we have set $k=1$. In this background,
 we have $R_{ij} = 2 \gamma_{ij}/a^{2} $ and $R = 6/a^{2}$. 
The Hamiltonian constraint in this minisuperspace approximation takes the following form,  
$
[\pi_{a}^{2}-\omega^2(a)]=0, 
$
where 
$
 \omega^2(a)  = - 3\pi^2 (3a ^2 - \Lambda a^4 )/4G
$.
Now we promote the canonical momentum $\pi_{a}$ to an operator, and so we have 
$\pi_{a}=-i\partial_{a}$.
Thus, the Wheeler-DeWitt equation corresponding 
to this classical Hamiltonian constraint is given by
\begin{equation}
\left[\frac{\partial^{2}}{\partial a^{2}}+ \omega^2(a)  \right]\phi [a] =0. \label{a}
\end{equation}
We interpret  Eq. $(\ref{a})$, 
as a classical field equation of a classical 
field $\phi(a)$, whose classical  action is given by 
\begin{equation}
\mathcal{S} [\phi (a)] = \frac{1}{2}
\int  da \left( \left({\frac{\partial \phi}{\partial a}}\right)^2 
- \omega^2(a) \phi^2 \right). \label{t}
\end{equation}
Now obviously the variation of this action for third quantization 
 $\mathcal{S}[\phi (a)] $, given by Eq. $(\ref{t})$,
 will lead to the Wheeler-DeWitt equation $(\ref{a})$. 
If we  interpret  the scaling factor  $a$ as the time variable, then
 the momentum conjugate to $\phi(a)$ will be $ p = \dot{\phi}$, 
 where   $ \dot{\phi} $ is the derivative of $\phi$ with respect to $a$.   
Now the third quantized  Hamiltonian obtained by 
the Legendre transformation,   can be written as 
\begin{equation}
\mathcal{H} (\phi (a), p (a) )  = \frac{1}{2} p^2 + \frac{\omega^2(a)}{2}
\phi^2 . \label{c}
\end{equation}
This  Hamiltonian given by Eq. $(\ref{c})$  is the  Hamiltonian for the harmonic oscillator with
time-dependent frequency $\omega(a)$ \cite{td, zl, lz,  z1}. 
We first third quantize this Hamiltonian by  imposing equal
time commutation relation between $\phi (a)$ and $p(a)$, $[\phi (a), p(a)] =i$. 
Then we go to the Schroedinger's picture and diagonalize the operator version of $\phi (a)$. 
In the Schroedinger's picture this diagonalized  operator version of $\phi (a)$ is  represented 
by a time independent field $\phi$, and the operator version of $p(a)$ is represented by $- i\partial/\partial \phi$. 
Now we can write the third quantized Schroedinger equation for the multiverse as \cite{1q}
\begin{eqnarray}
 O_0(a, \phi ) \Phi     (\phi, a) &=& \left[ - i\frac{\partial}{\partial a }  +
 \mathcal{H} \left(\phi , p  \right)  \right]\Phi(\phi ,a)
 \nonumber \\ &=& \left[ - i\frac{\partial}{\partial a }  - \frac{1}{2}\frac{\partial^2 }{\partial \phi^2 } + 
\frac{\omega^2(a)}{2} \phi^2 \right]\Phi(\phi ,a)\nonumber \\ &=& 0. \label{thir}
\end{eqnarray}
Here $\Phi(\phi ,a)$ is the wave function for the multiverse, 
and different modes correspond to different universes.  

If we denote the two linearly independent solutions to 
the third quantized Schroedinger  as $u_0(a)$ and $v_0(a)$,
and define $\rho_0(a)=\sqrt{u_0^2+v_0^2}$, then we have  \cite{td1}
\begin{eqnarray}
{d^2 \over da^2}{\rho_0}+\omega^2(a)\rho_0 - {\Omega_0^2 \over \rho_0^3}=0, 
\end{eqnarray}
where $\Omega_0$ is  given by $\Omega_0 = \dot{v}_0u_0-\dot{u}_0v_0$.
Let the third  quantized wave functions $\Phi ^s(\phi ,b)$ be the solution to the following equation, 
\begin{eqnarray}
 O_s(b,\phi) \Phi^s (\phi,b) &=& \left[ - i\frac{\partial}{\partial b } 
 + \mathcal{H}^s \left(\phi, p  \right)  \right]\Phi ^s(\phi ,b)
 \nonumber \\ &=& \left[ - i\frac{\partial}{\partial b }  - \frac{1}{2}\frac{\partial^2 }{\partial \phi^2 } + 
\frac{\phi^2 }{2}  \right]\Phi^s(\phi ,b)\nonumber \\ &=& 0. 
\end{eqnarray} 
This is the equation of a  simple harmonic oscillator system of unit mass and frequency. Here $b$ denotes the time 
and  $\Phi ^s(\phi ,b )$ denote  wave functions of this simple harmonic oscillator. 
This time  $b$ is related to $a $  through the relation, 
$
\rho_0^2 d b =\Omega_0 da.
$
Now we define   $U_{\omega}(\rho_0,\Omega_0)$ as
\begin{eqnarray}
U_{\omega}(\rho_0,\Omega_0)=\exp\left({i\dot{\rho_0} \over 2 \rho_0}\phi ^2\right)
  \exp\left[-{i\over 2}\left(\ln{\rho_0\over \sqrt{\Omega_0}}\right)\left(\phi p+p\phi \right)\right],
\end{eqnarray}
where 
$
{ \Omega_0}U_{\omega}O_s(b, \phi )U_{\omega}^\dagger\mid_{b =b (a)}= \rho_0^2 O_0(a, \phi), 
$
and 
$
\Phi    (\phi ,a)=U_{\omega}\Phi ^s|_{b =b (a)}.
$
Furthermore, the wave function for the simple harmonic oscillator can be written as  
\begin{eqnarray}
\Phi ^s(\phi ,b )|_{b =b (a)}&=&
          {1\over \sqrt{2^nn!\sqrt{\dot{\Phi}}}} e^{-i(n+{1\over 2})b } 
  \exp\left[-{\phi ^2\over 2}\right]H_n(   \phi )|_{b =b (a)}\nonumber \\
&=&{1\over \sqrt{2^nn!\sqrt{\dot{\Phi}  }}}
          \left({u_0(t)-iv_0(a) \over \rho_0(a)}\right)^{n+1/2}  \nonumber \\ && 
\times \exp\left[-{\phi ^2\over 2  }\right]H_n(   \phi ).
\end{eqnarray}

We also define  $\delta_{u_1}(a)$ as 
$
2\dot{\delta}_{u_1}=\omega^2u_1^2-\dot{u}_1^2
$
where $u_1$ is a linear combination of $u_0(a)$ and $v_0(a)$. Furthermore, 
  $U_f$ can be written as 
$
U_f=\exp[{i}(\dot{u}_1\phi +\delta_{u_1}(a)]\exp(-{i }u_1p),
$
where 
$
U_fO_0U_f^\dagger=O_0.
$
Therefore, the wave functions $\Phi  (\phi ,a)  $ of the multiverse is given by \cite{td1aa}
\begin{eqnarray}
\Phi(\phi ,a)&=&U_fU_{\omega}\Phi ^s(\phi,b )\mid_{b =b (a)}\cr
&=&{1\over \sqrt{2^nn!\rho_0(a)}}\left({\Omega_0\over \pi  }\right)^{1/4}
\left({u_0(t)-iv_0(t) \over \rho_0(a)}\right)^{n+1/2}\cr
&&\times \exp\left[{i}(\dot{u}_1(a)\phi +\delta_{u_1}(a))\right]H_n\left(\sqrt{\Omega_0   }{\phi -u_1(a) \over \rho_0(a)}\right)
\cr&&\times
\exp\left[{(\phi -u_1(a))^2\over 2  }
\left(-{\Omega_0\over \rho_0^2(a)}+i{\dot{\rho}_0 \over \rho_0}\right)\right].
\end{eqnarray}  
Thus, we have obtained the wave function for the full multiverse. The different modes here correspond to different universes. 

\section{Fourth Quantization}
In the previous section we obtained the third quantized Schroedinger's equation for the multiverse. 
Now we can repeat the procedure of going from first quantization to second quantization  or 
 going from second quantization to third quantization, to go from third quantization to fourth 
 quantization. In order to do that we can view the  third quantized Schroedinger's equation for the multiverse, 
 as a classical field equation and construct Lagrangian which generates this 
 classical field equation. We can use this Lagrangian to calculate the momentum conjugate to this field variable. 
 Then we can use it to fourth quantize this theory and construct a fourth quantized Fock space. 
 In this Fock space the creation and annihilation operators will create and annihilate multiverses.  
 Thus, we will obtain a theory of multi-multiverses.  

So, to analyse the fourth quantization of the Wheeler-DeWitt equation,  
 we will view Eq.  $(\ref{thir})$ as a classical field equation. 
In Eq.  $(\ref{thir})$, $\phi$ acts like the space 
coordinate,  $a$ acts like the  time coordinate,  and $\omega^2(a) \phi^2$ acts like a spacetime 
dependent mass. Thus, this equation  looks like the classical field equation for a quantum field theory in two dimensions 
with a spacetime dependent mass. 
So,  we define a fourth quantized action  for it as 
\begin{equation}
S = \int d\phi da   \Phi (\phi, a) O_0(a, \phi ) \Phi     (\phi, a).
\end{equation}
It may be noted that instead of a Schroedinger's like equation for the third quantized multiverse, we can 
 use a Klein-Gordon  like equation for third quantized multiverse. 
 For a  Klein-Gordon  like equation for third quantized multiverse
 the time dependence of 
 $O_0(a, \phi )$ will be given by $\partial^2 /\partial a^2 $. 
Now we define the following  symplectic product 
\begin{equation}
 (\Phi, \Phi') = -i \int d \phi  [\Phi^* (\phi , a)\dot{\Phi}' (\phi , a)- \Phi'(\phi , a) \dot{\Phi}^{*}(\phi , a)]. 
\label{hermihermi}
\end{equation}
We let $\{\Phi_n\}$ and $\{\Phi^*_n\}$ form a complete set of solutions to the third quantized  
equation 
for the multiverse, and suppose 
\begin{eqnarray}
 (\Phi_n,\Phi_m) &=& M_{nm}, \label{conditions1}\\ 
(\Phi_n,\Phi^*_m) &=& 0, \label{conditions2}\\ 
(\Phi^*_n,\Phi^*_m) &=& - M_{nm}.\label{conditions3}
\end{eqnarray}
The condition given in Eq. $(\ref{conditions2})$ does not hold in general and is  imposed here 
as a requirement on the 
complete set of solutions to the third quantized  equation for the multiverse. 
We will also require $M_{nm}$ to only have positive eigenvalues as this does not also hold in general. 
This  matrix $M_{nm}$ is  Hermitian because 
\begin{eqnarray} 
 M_{nm} &=& -i \int d \phi  [ \Phi_n^*(\phi , a)\dot{\Phi}_m(\phi , a) - 
 \Phi_m(\phi , a) \dot{\Phi}_n^{*}(\phi , a)] \nonumber \\ 
&=&\left[ -i  \int d \phi   [ \Phi_m^* (\phi , a)\dot{\Phi}_n (\phi , a)- 
\Phi_n (\phi , a)\dot{\Phi}_m^{*}(\phi , a)] \right]^* \nonumber \\ 
&=& M^*_{mn }.\label{bov21}
\end{eqnarray}

Now in analogy with second quantized  quantum field theory, we promote 
 $\Phi$ and $\dot{\Phi}$  to  Hermitian operators, 
and  impose the following, 
\begin{eqnarray}
 \left[     {\Phi}(\phi ,a ),     {\dot{\Phi}} (\phi ', a)\right]  &=& i \delta (\phi ,\phi '), \nonumber \\ 
  \left[      {\Phi}(\phi , a),     {\Phi} (\phi ', a)\right]  & =& 0,\nonumber \\  
\left[     {\dot{\Phi}}(\phi , a),     {\dot{\Phi}} (\phi ', a)\right]  &=& 0,\label{ccr3}
\end{eqnarray}
Now we can express $    {\Phi}(\phi, a) $  as, 
 \begin{equation}
      {\Phi} (\phi, a) = \sum_n [a_n \Phi_n + a^{\dagger}_n \Phi^*_n ]. \label{bov}
 \end{equation}
because $\{\Phi_n\}$ and $\{\Phi^*_n\}$ form a complete  set of solutions to the third quantized  
equation for the multiverse. 
 We define the fourth quantized vacuum state $|0\rangle$, as the state  that  is annihilated by $a_n$,
$
  a_n |0\rangle = 0
$. This state  is a purely mathematical object. It contains neither matter nor spacetime. 
However, we can create matter and spacetime from this vacuum state using $a_n^{\dagger}$. Thus, 
multiverse can be built  from this  vacuum state.

It may be noted that  the division between $\{\Phi_n\}$ and $\{\Phi^*_n\}$ is not unique even after imposing conditions given by
 Eqs. $(\ref{conditions1})$-$(\ref{conditions3})$. Due to this non-uniqueness in division between
 $\{\Phi_n\}$ and $\{\Phi^*_n\}$, there is   non-uniqueness in the definition of the fourth quantized vacuum state also. 
This can be seen by considering  $\{\Phi'_n\}$ and $\{{\Phi'_m}^*\}$ as another  
 complete set of solutions to the third quantized  equation  for the multiverse. 
Now we  
express ${\Phi} (\phi, a)$ as
\begin{equation}
       {\Phi} (\phi, a) = \sum_n [a'_n \Phi'_n + a'^{\dagger}_n \Phi'^*_n]. \label{bov1}
\end{equation}
Here $\{\Phi'_n\}$ and $\{{\Phi'_m}^*\}$ also satisfying
 conditions similar to the conditions given in  Eqs. $(\ref{conditions1})$-$(\ref{conditions3})$.
We define a corresponding  fourth  quantized vacuum state $|0'\rangle$ as   the state annihilated by $a'_n$,
$
 a'_n|0'\rangle =0
$.
The spacetime and matter  can again  be built  by repeated action of $a'^{\dagger}_n$ on $|0'\rangle$.
As $\Phi_n$ and $\Phi^*_n$ form a complete set of  solutions to the third quantized  equation for the 
multiverse, we can  express $\Phi'_n$  
as a linear combination of $\Phi_n$ and $\Phi^*_n$, 
\begin{equation}
 \Phi'_n = \sum_m [\alpha_{nm}\Phi_m + \beta_{nm}\Phi^*_m].\label{bovv2} 
\end{equation}
Thus, we have 
\begin{eqnarray}
 a_n &=& \sum_m [\alpha_{nm}a'_m + \beta^*_{nm}a'^{\dagger}_m],\\
a^{\dagger}_n &=& \sum_m [\alpha^*_{nm} a'^{\dagger}_m + \beta_{nm}a'_m].
\end{eqnarray}
 As long  
 as $\beta_{nm} \neq 0$,  these two Fock spaces based on different complete set of  
 solutions to third quantized  equation for 
the multiverse are different. So,  $
 a_n |0\rangle =0 $, however 
\begin{eqnarray}
  a_n|0'\rangle &=& \sum_m [\alpha_{nm}a'_m + \beta^*_{nm}a'^{\dagger}_m]|0'\rangle  \nonumber \\
&=&\sum_m \beta^*_{nm}a'^{\dagger}_m |0'\rangle \neq 0.
\end{eqnarray}
So, $a_n|0'\rangle$ is  a multiverse state given by 
\begin{equation}
\langle 0'| a_n^{\dagger} a_n|0'\rangle = \sum_m \sum_k \beta_{nm}\beta^*_{nk}M_{mk}.
\end{equation}
Thus, even though we started from a vacuum state, we ended up with the multiverse. 

\section{Interactions }

It is well know that in a  second quantized theory, the  interaction terms in the Lagrangian create and annihilate particles. 
Furthermore, in a third quantized theory, the  interaction terms in the Lagrangian create and annihilate universes \cite{b, k}.
Similarly, in a fourth quantized theory, the  interaction terms in the Lagrangian create and annihilate multiverses. 
So, in this section we will
 analyse the effect of interaction for the multi-multiverses. 
 
To analyse the interaction of these multiverses,  we first start from the free fourth quantized action
\begin{equation}
S = \int d\phi da   \Phi (\phi, a) O_0(a, \phi ) \Phi     (\phi, a),
\end{equation}
and then we can add interaction terms $S_I[\Phi (\phi, a)]$ to it,
\begin{equation}
S_{T} [\Phi (\phi, a)] =  S [\Phi (\phi, a)]
+ S_I[\Phi (\phi, a)].
\end{equation}
In order to analyse the perturbation theory of these multiverses, we first  define 
\begin{equation}
 J\Phi = \int d\phi da J(\phi, a) \Phi (\phi, a)
\end{equation}
 and then let  $a \to ia$.
Now we can write the Euclidean partition function as
\begin{equation}
 Z [J] = \int D\Phi \,\exp - \mathcal{S} [\Phi ]_{ET} + J\Phi ,
\end{equation}
where $\mathcal{S}[\Phi]_{ET}$ is the Euclidean version of $\mathcal{S} [\Phi]_{T}$ obtained by letting $a \to i a$. 
Now we can define 
\begin{equation}
 W[J] = - \ln Z[J].
\end{equation}
We can also define the effective action as 
\begin{equation}
 \Gamma [\Phi_b ] = W[J] - \Phi_b J, 
\end{equation}
where 
\begin{equation}
 \Phi_b = \frac{\delta W[J]}{\delta J}.
\end{equation}
The full quantum equation of motion will be given by 
\begin{equation}
 \frac{\delta \Gamma}{\delta \Phi_b} =0. 
\end{equation}

We will now analyse the formation of two multiverses,   $M_3$ and $M_4$, from 
the  collision of two other multiverses,  $M_1$ and $M_2$.
  The interaction term  which generates this process is given by
\begin{equation}
  \mathcal{S}_{EI} = \frac{\lambda }{4!}
\int d \phi  da  \Phi^4(\phi,a).
\end{equation}
If we represent matter and gauge fields collectively by $\chi$
 and include the contribution  from $\chi$ in our formalism,
then $\phi$ would also depend on $\chi$, 
and so, $\Phi $ would also depend on $\chi$.  
Let $B$ be  a third quantized charge  that remains conserved when two 
universes collide \cite{thir}.
Also let the total number of universes (with a positive value of $B$ )
 and anti-universes (with a negative value of $B$ )  in multiverse $M_k$ 
( with $k =1,2,3,4$) be $n_k $ and $m_k$, respectively.  Now if   the multiverse $M_1$
and $M_2$
 have formed from nothing without violating the   conservation of $B$, then we have, 
$
 Bn_1 - Bm_1 + Bn_2 -Bm_2 =0.
$
Here $Bn_1 + Bn_2$  represent the total $B$ number of the universes
  and $-Bm_1 -Bm_2$ represent the total $B$ number of the  anti-universes. 
Furthermore, after the collision, if the $B$ number is conserved, we will have, 
$
 Bn_3 - Bm_3 + Bn_4  - Bm_4 =0. 
$
However the  $B$ number in the multiverse 
$M_3$ or $M_4$ need not be separately conserved, so we have,    
$
 Bn_3 - Bm_3 \neq 0, $ and $ 
 Bn_4 - Bm_4 \neq 0
$.
Thus, after the collision some multiverse can have more positive 
$B$ than  if the other multiverse has more negative $B$.  
Previously, it was proposed that an initial universe broke 
into two universes and this was proposed as a explanation 
for the dominance of matter over anti-matter in our universe \cite{b}. 
In this paper this model is generalized to include the full multiverse.

\section{Conclusion}
In this paper we analysed third quantization of Wheeler-Dewitt equation for 
a Friedman-Robertson-Walker universe 
with a cosmological constant. We analyse the multiverse in this formalism and calculated 
the wave function  of the multiverse. 
We  also constructed the  Fock space for the multi-multiverses.  As there was  no single   vacuum state,
 we can ended up with a multiverse state, even after starting from a vacuum state.
 This was used as an possible explanation of the 
creation of multiverse. We also analysed the effect of interactions for the multi-multiverses.  

It is known  that in second quantized quantum field theory, we can construct 
 conserved charges which remain conserved, 
even when the particle number is not conserved. These conserved charges are 
generated by the invariance of the second quantized 
Lagrangian under some symmetry. In analogy with second quantized quantum field theory, 
we can also construct a third quantized 
Noether's theorem. Thus, if there is a symmetry which will leave a third quantized 
Lagrangian invariant, then the charge corresponding 
to it will be conserved even if the number of universes is not conserved.
  Here we  argued that a third quantized Noether's charge will not remain conserved 
for individual multiverse, in case of a fourth quantized interacting theory. It 
will be interesting to construct  third quantized Noether's charges for different 
quantum cosmological models and analyse the conservation of these Noether's charges
in a fourth quantized theory.

It may be noted that canonical quantum gravity has evolved into loop quantum gravity \cite{p,q,r,s}. 
Third quantization of loop quantum gravity has led to the development of group field theory \cite{q1a,s1,r1,t1}. 
Recently, group field cosmology has also been developed \cite{gf,  gf2}. 
It will be interesting to analyse the creation of multiverse from a vacuum state in group field cosmology. Furthermore, 
the group field cosmology has also been supersymmetrized \cite{gf1}.
 In this supergroup field cosmology, 
there are universes with both bosonic and fermionic distributions in the multiverse. It is hoped that by considering 
a supersymmetric distribution of universes, we might get better understanding of the cosmological constant. 
Thus, it will be interesting to include supersymmetry in  this present work 
and analyse multi-multiverses  
containing both bosonic and fermionic multiverses.

\end{document}